\documentclass[aps, prc, reprint, amsmath, groupedaddress, nofootinbib]{revtex4-1}
\usepackage[utf8]{inputenc}
\usepackage{hyperref}
\usepackage{amsmath}
\usepackage{amssymb}
\usepackage{amsfonts}
\usepackage{tabularx}
\usepackage{booktabs}
\usepackage{graphicx}
\usepackage{color}
\usepackage{multirow}
\usepackage{verbatim}
\usepackage[inline]{enumitem}
\graphicspath{{fig/}}
\definecolor{theblue}{RGB}{0,50,230}
\usepackage{appendix}
\hypersetup{
  colorlinks=true,
  linkcolor=theblue,
  citecolor=theblue,
  urlcolor=theblue
}

\newcommand{\trento}{T\raisebox{-0.5ex}{R}ENTo}

\newcommand{\Kpara}{\kappa_{\|}}
\newcommand{\Kperp}{\kappa_{\perp}}

\begin{abstract}
In relativistic heavy-ion collisions, the production of heavy quarks at large transverse momenta is strongly suppressed compared to proton-proton collisions. In addition an unexpectedly large azimuthal anisotropy was observed for the emission of charmed hadrons in non-central collisions. 
Both observations pose challenges to the theoretical understanding of the coupling between heavy quarks and the quark-gluon plasma produced in these collisions. Transport models for the evolution of heavy quarks in a QCD medium offer the opportunity to study these effects - two of the most successful approaches are based on the linearized Boltzmann transport equation and the Langevin equation.
In this work, we develop a hybrid transport model that combines the strengths of both of these approaches:
heavy quarks scatter with medium partons using matrix-elements calculated in perturbative QCD, while between these discrete hard scatterings they evolve using a Langevin equation with empirical transport coefficients to capture the non-perturbative soft part of the interaction.
With the hybrid transport model coupled to a state-of-the-art event-by-event bulk evolution model based on 2+1D relativistic viscous fluid dynamics, we study the azimuthal anisotropy and nuclear modification factor of heavy quarks in Pb+Pb collisions at $\sqrt{s} = 5.02$ TeV.
The parameters of our model are calibrated using a Bayesian analysis comparing to available $D$-meson and $B$-meson data at the LHC.
Using the calibrated model, we study the implications on heavy-flavor transport properties and predict novel observables.
\end{abstract}

\begin{document}
\title{A linearized Boltzmann -- Langevin model for heavy quark transport in hot and dense QCD matter}
\author{Weiyao Ke}
\author{Yingru Xu}
\author{Steffen A.\ Bass}
\affiliation{Department of Physics, Duke University, Durham, NC 27708-0305}
\date{\today}
\maketitle

\section{Introduction}
In recent years, the Relativistic Heavy-Ion Collider (RHIC) at Brookhaven National Laboratory and subsequently the Large Hadron Collider (LHC) at CERN have discovered a new state of matter in ultra-relativistic heavy-ion collisions, referred to as the quark-gluon plasma (QGP).
The three key observations that lead to 
the discovery of the QGP are the measured strong collective flow of bulk matter, parton recombination as manifest in constituent quark number scaling laws and jet energy-loss (i.e. jet quenching). 

The observed collective flow reveals that the bulk medium of the QGP undergoes a strong collective expansion after its initial creation.
This behavior can be explained in surprisingly great detail by models using relativistic viscous hydrodynamics.
Jet quenching refers to the strong suppression of the yield of high transverse momentum hadrons in nuclear collisions, compared to the scaled yield in proton-proton collisions where medium effects are assumed to be small.
Calculations have shown that this suppression is a consequence of jets losing energy to the hot, dense and color-deconfined medium. 

Heavy quarks (charm and bottom) are often seen as complementary probes of the QGP, but partly also belong to the category of jet observables, depending on their transverse momenta. Their large masses (compared to the prevailing temperatures generated in collisions at current heavy-ion colliders) constrain their production to early reaction times via hard perturbative Quantum-Chromodynamics (pQCD) processes. Flavor conservation ensures that the overwhelming majority of heavy quarks survive the entire reaction, allowing them to probe the full space-time evolution of the reaction.
These two features are particular attractive to theorists as these flavor-tagged particles are much easier to track in the calculations than the evolution of a full jet.
The mass also sets an additional energy scale to the problem and brings rich physics to the heavy-flavor sector.
In the high transverse momentum region, heavy quarks lose energy mainly through radiative processes connecting them to jet energy loss studies \cite{Wicks:2007am, Djordjevic:2004nq, Xu:2014tda, Kang:2016ofv}, whereas
in the low transverse momentum region their large mass delays their thermalization, providing a window to study the equilibration process \cite{Moore:2004tg,Riek:2010fk,Cao:2013ita}.
Heavy flavors are therefore ideal and unique probes to determine QGP properties.

The in-medium propagation of heavy quarks is often studied in a kinetic approach that is linearized with respect to the heavy quark distribution function and the medium particle distribution function is assumed to be thermal, obtained from hydrodynamic models.
The linearization implies that any effects of the heavy quark interactions on the medium are neglected.
The linearized Boltzmann transport equation and the Langevin equation are both widely used linearized models but make different assumptions regarding the nature of the interaction and thus often focus on different regimes in the heavy quark phase space \cite{Auvinen:2009qm,Cao:2016gvr, Cao:2017hhk, PhysRevD.37.2484, Moore:2004tg}.
The linearized Boltzmann transport equation is based on elementary scattering processes that can be directly calculated, e.g. via pQCD.
However, calculations in the presence of a medium are extremely complicated even at leading order \cite{Arnold:2002zm}.
Also, the pQCD processes are often plagued by soft divergences that need to be regulated by a medium scale proportional to temperature. Moreover, at current collision energies the relevant temperature is not high enough which creates ambiguities for the pQCD calculation through the scale dependence of the strong coupling constant $\alpha_s$.

The Langevin equation takes a different approach: 
it assumes that heavy quark receives frequent but soft momentum kicks from the medium, making a statistical description of the interaction possible -- in terms of ``drag" and ``diffusion" coefficients.
These transport coefficients encode the first and second moments of the momentum-exchange rate but are agnostic to further details of the elementary processes and medium properties.
There are efforts to calculate these transport coefficients in various approaches including lattice QCD \cite{Moore:2004tg,CaronHuot:2008uh, Gossiaux:2008jv,He:2012df,Riek:2010fk,vanHees:2007me,Scardina:2017ipo,Ding:2012sp,Banerjee:2011ra,Francis:2015daa}. Our group has taken a complementary approach, using experiment data to calibrate our Langevin based transport model to measured observables and thus extract the transport coefficients directly from data via a Bayesian analysis \cite{Xu:2017obm}. The drawback of this approach is that it does not in itself provide a fundamental understanding of the interaction mechanism but can only provide guidance to direct calculations of the transport coefficients in terms of compatibility to experimental observation.

In this work, we propose to combine the strengths of the linearized Boltzmann equation approach with that of the Langevin equation to develop a hybrid transport model for the evolution of heavy quarks in a QGP medium.
In this hybrid model, called {\tt Lido} ({\bf Li}nearized Boltzmann with {\bf d}iffusion m{\bf o}del),  the heavy quarks scatter off medium particles described by a linearized Boltzmann equation with pQCD matrix elements (the scattering component), and between scatterings propagate according to a Langevin equation (the diffusion component) with empirical temperature- and momentum-dependent transport coefficients to describe the soft non-perturbative components of the interaction missing from the above scattering picture.
Both elastic and inelastic scatterings are included in the scattering component with the soft divergence screened by a Debye mass $m_D$ and the Landau-Pomeranchuk-Migdal (LPM) effect taken into account effectively.
The scattering process inside a medium is a multi-scale problem that includes a momentum transfer scale $Q$ and a medium scale that is proportional to the temperature $\mu\pi T$.
The QCD coupling constant has a scale dependence that we choose to be the maximum of $Q$ and $\mu\pi T$, which means the typical scale of an in-medium process is cut off by the medium scale.
The details of the running coupling constant we have utilized can be found in Appendix \ref{appendix:alphas}.
The medium scale parameter $\mu$ is the only parameter in the scattering component and we assume it encodes the uncertainty in the pQCD matrix-element approach in our models.
The diffusion component has several parameters depending on the way in which transport coefficients are parametrized. 
The idea is to include non-perturbative contributions in terms of these transport coefficients.
For future studies, we will also consider absorbing small-momentum-transfer elastic pQCD scatterings and the associated radiation into a radiation-improved Langevin equation component of the model.
We would like to point out that a rigorous separation of matrix-element based scattering and diffusion has been proposed for the study of light parton jet energy loss up to next-to-leading order in pQCD \cite{Ghiglieri:2015ala}.
In our study, we don't require the diffusion component to be perturbative in nature.

All parameters of the model will be calibrated to data using Bayesian inference \cite{Novak:2013bqa,Bernhard:2015hxa}.
This approach takes experimental uncertainties into account and provides the probability distributions for all model parameters given the experimental data.
The Bayesian technique is particularly useful for focusing on a subset of parameters such as the transport coefficients.
It allows the marginalization over all other parameters and computes the probability distribution for the parameters of interest. 
The marginalization provides a parameter range that is not only preferred by the experiments, but also already includes uncertainties in the other model parameters.
Therefore the Bayesian technique reveals what actually can be learned from the data, considering both experimental accuracy and model uncertainties.
The Bayesian methodology has been successfully applied to the extraction of initial condition and bulk transport coefficients of the soft QGP medium \cite{Novak:2013bqa, Pratt:2015zsa, Bernhard:2015hxa, Bernhard:2016tnd, Auvinen:2017fjw} and to the heavy quark sector for the extraction of the heavy quark momentum diffusion parameter $\hat{q}$ using a radiation improved Langevin equation \cite{Xu:2017obm, Cao:2013ita}.
In this work, we shall perform a likewise extraction of the heavy quark transport properties using the proposed {\tt Lido} model and compare with previous calculations to see how the results dependent on the use of different transport approaches.

The paper is organized as follows. 
We describe the model in detail in section \ref{section:model}. In section \ref{section:test}, the model is tested in a static medium with a set of default parameters. 
We calibrate the model parameters in section \ref{section:calibration} to data and predict novel observables using high likelihood parameter values in section \ref{section:prediction}. 
Finally, section \ref{section:conclusion} contains summary and discussion of results.

\section{Heavy quark propagation in a hybrid transport model}\label{section:model}
As introduced in the previous section, the {\tt Lido} model consists of a linearized-Boltzmann equation of scatterings $\mathcal{C}[f_Q]$ and a diffusion component that appears as a Fokker-Planck operator $\mathcal{D}[f_Q]$ in the transport equation:
\begin{eqnarray}
\nonumber
  \frac{p \cdot \partial f_Q}{E}  &=& 
\mathcal{C}[f_Q] - \frac{\partial}{\partial p_i}\left(A_i -\frac{1}{2}\frac{\partial}{\partial p_j}B_{ij} \right) f_Q \\
  &=& 
\left( \mathcal{\hat{C}} + \mathcal{\hat{D}} \right) f_Q,
\end{eqnarray}
with a formal solution,
\begin{eqnarray}
\nonumber
f_Q(x,p) &=& \exp\left\{ \int_{x'}^x \gamma u \cdot dx \left( \mathcal{\hat{C}} + \mathcal{\hat{D}} \right) \right\} f_Q(x',p)\\
&\approx & e^{\Delta t \hat{C}}e^{\Delta t \hat{D}} f_Q(x', p) + \mathcal{O}(\Delta t^2)
\end{eqnarray}
Technically, this can be solved by the split step method within a tiny time step $\Delta t = \gamma u \cdot dx$, during which we apply the operation of scattering and diffusion subsequently up to corrections of $\mathcal{O}(\Delta t^2)$.
Next, we discuss the physics included in each components in detail. 

\subsection{Scattering component}
The scattering of heavy quarks with medium particles is treated with a linearized Boltzmann equation,
\begin{eqnarray}
    \frac{p \cdot \partial f_Q}{E} = \mathcal{C}[f_Q].
\end{eqnarray}
The left hand side represents the free evolution of the heavy quark distribution function. 
Scatterings with medium partons modify the distribution function via the collision integral on the right.
The medium partons are assumed to obey classical statistics for simplicity, whose thermal occupancy number follows the Maxwell--J\"uttner distribution, 
\begin{eqnarray}
f_{q,\bar{q}, g}(t,x,p) = \exp\left(-\frac{p \cdot u(t,x)}{T(t,x)}\right),
\end{eqnarray}
and any non-equilibrium corrections are neglected.
The space-time evolution of the temperature field $T$ and velocity field $u^\mu$ are obtained in an event-by-event 2+1D viscous relativistic hydrodynamic calculation \cite{Heinz:2005bw,Song:2007ux,Shen:2014vra}.
We solve $f_Q(t,x,p)$ using Monte Carlo techniques by representing the distribution function with an ensemble of heavy quarks.
Within a given time step $\Delta t$, each heavy quark scatters according to its reaction probability $\Delta P$.
We always calculate $\Delta P$ in the rest frame of the fluid cell with given temperature $T$ and velocity field $u^\mu$.
In this reference frame, the heavy quark with energy $E_1$ can collide with medium particles that together form an $n$-body initial state denoted as $\{\textrm{in}\}$, and the outgoing particles after the collision form the $m$-body final state $\{\textrm{out}\}$.
The probability for a heavy quark to undergo an interaction of a certain type inside the fluid cell per unit time is the so called scattering rate $\Gamma$,
\begin{eqnarray}\label{eq:rate}
    \frac{dP}{dt} &=& \Gamma(E_1, T, t) \nonumber \\
    &=& \frac{d}{\nu} \frac{(2\pi)^3\delta}{\delta f_Q(p_1)}\int d\Phi(n,m) \prod_{\textrm{\{in\}}} f_i(p_i) 
\overline{|M|^2},
\end{eqnarray}
where the $d\Phi(n,m)$ is the $(n+m)$-body phase-space integration,
\begin{eqnarray}
\nonumber
d\Phi(n,m) = (2\pi)^4\delta^4\left(P_{\textrm{in}}-P_{\textrm{out}}\right)\prod_{\{\textrm{in, out}\}} \frac{dp_i^3}{2E_i(2\pi)^3} 
\end{eqnarray}
$\overline{|M|^2}$ is the initial state spin-color averaged scattering matrix-element squared.
The factor $d$ denotes the degeneracy of the incoming medium particles and $\nu$ is the symmetry factor of identical particles in the initial / final state of the collision.
If a scattering process occurs within $\Delta t$ according to the probability $\Delta P=\Gamma \Delta t$, the details of the initial and final states can be obtained by sampling the differential scattering rate over the $(m+n-1)$-body phase space.
The many-body phase-space sampling may look formidable at first sight, but can be factorized into sequential initial-state and final-state sampling as long as one uses classical statistics and a simple version of the medium screening effect.
The relevant sampling details can be found in Appendix \ref{appendix:sample}.
The time step $\Delta t$ is chosen small enough so that the probability of multiple scatterings is negligible. 

\begin{figure}
\includegraphics[width=.5\textwidth]{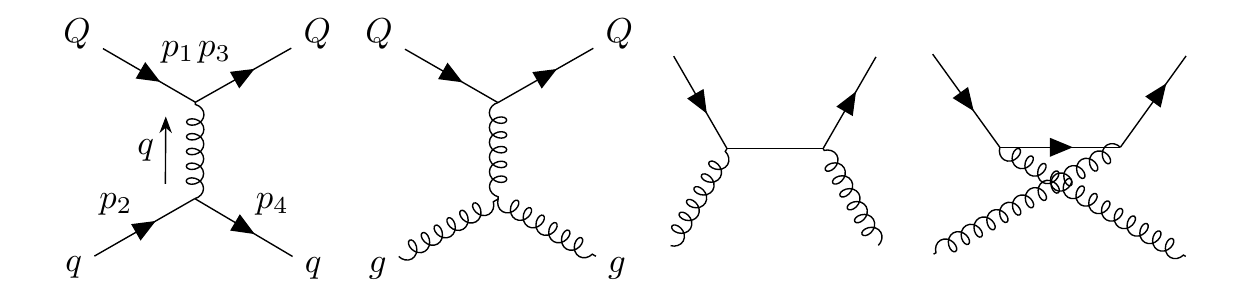}
\caption{Elastic processes: The first diagram corresponds to heavy quark ($Q$) - light quark ($q$, $\bar{q}$) scattering. The last three diagrams contribute to heavy quark ($Q$) - gluon ($g$) scattering.}\label{plots:feyn-elastic}
\end{figure}

Focusing on the processes to be included in the collision term, 
it has been shown that at leading order, an energetic parton can scatter elastically with a medium parton (light quark and anti-quark or gluon) or emit a gluon triggered by multiple soft collisions which we call an inelastic process based on its particle number changing nature \cite{Arnold:2002zm}.
We will keep using the terms ``elastic" and ``inelastic" to distinguish between these two types of processes and their associated energy loss.
For elastic processes, quark-gluon scattering contributes three diagrams corresponding to $s, t,$ and $u$ channel momentum exchange; quark-quark scattering only has a $t$ channel contribution. 
These diagrams are shown in Fig. \ref{plots:feyn-elastic} and the matrix-elements for these processes in vacuum are available at leading order in pQCD (see \ref{appendix:matrix-element}).
In these expressions, the characteristic $t-$channel gluon propagator causes a divergence in the cross-section as the momentum transfer vanishes,
\begin{eqnarray}
d\sigma \propto \frac{1}{t^2} dt.
\end{eqnarray}
In a quark-gluon plasma medium, those soft gluon excitations constantly interact with thermal particles causing this divergence to be screened by a Debye mass in the static limit \cite{Moore:2004tg}.
Generally, the gluon propagator should be replaced by a hard-thermal loop (HTL) propagator \cite{Peshier:1998dy}.
Using a HTL propagator involves a complicated self energy that depends on the medium reference frame, making it hard to implement in a cross-section based Monte-Carlo approach, where the calculation and sampling is easiest performed in the CoM frame of the collision.
Hence we choose to adopt the simple replacement $t^2 \rightarrow (t-\Lambda_{\textrm{QCD}}^2)(t - m_D^2)$ (up to an additional factor of $\Lambda_{QCD}^2$) that takes dynamic scattering center effects into account \cite{Djordjevic:2008iz}.

\begin{figure}
\includegraphics[width=.5\textwidth]{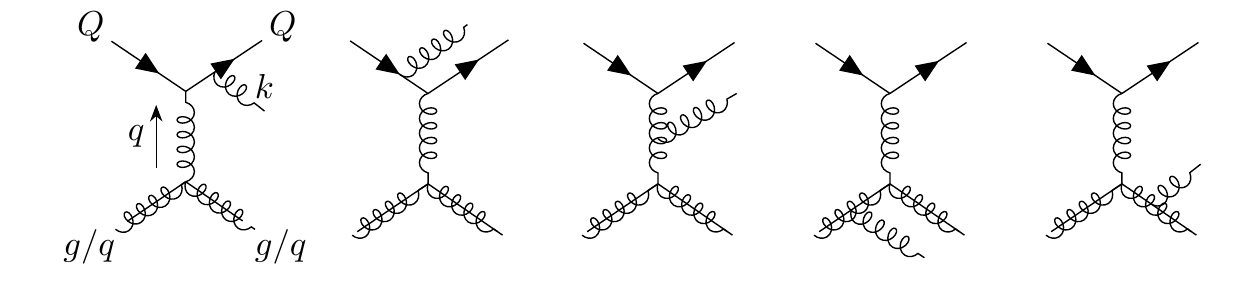}
\caption{Inelastic processes: a heavy quark collides with a medium light (anti-)quark or gluon and radiates an additional gluon.}\label{plots:feyn-inelastic}
\end{figure}

At large heavy quark energies, inelastic processes shown in Fig. \ref{plots:feyn-inelastic} become important.
We start from the case of gluon emission associated with the scattering with one medium particle, the effective treatment of multiple scatterings will be discussed later.
The corresponding matrix-elements are derived in an improved Gunion and Bertsch approximation in the high energy and soft gluon limit \cite{PhysRevD.25.746,Fochler:2013epa}, with the heavy quark mass effect (the dead-cone effect) included \cite{Uphoff:2014hza}.
Although the inelastic diagrams seem to involve one more power of $\alpha_s$, it actually contributes at leading order to the energy loss due to the small transverse momentum emission that we screened by the asymptotic gluon thermal mass $m_D/\sqrt{2}$ \cite{Ghiglieri:2015ala}.
The available phase space for radiating a gluon also grows with the heavy quark incident energy and it eventually becomes the dominant energy loss mechanism at high energies.
The Boltzmann equation including only the gluon radiation ($2\rightarrow 3$) process would violate detailed balance. We therefore include the reverse process namely gluon absorption ($3\rightarrow 2$) so that detailed balance is preserved and the model approaches the correct thermal equilibrium at large times.
The absorption process has been studied within the full Boltzmann equation \cite{Xu:2004mz} for light partons but so far has not been implement for heavy quarks and for any linearized transport models.
In the rate equation, the matrix-element for the absorption relates to the radiation matrix-element by sending the final state gluon to the initial state ($k^\mu \rightarrow -k^\mu$).
This initial state medium gluon brings an additional Boltzmann factor $e^{-k/T}$ to the differential rate.
For the fast moving heavy quark, most of the thermal gluons move in the opposite direction of the heavy quark in the center of mass frame,
so the probability for an energetic heavy quark to absorb a thermal gluon is extremely small. 
On the other hand, low energy heavy quarks can efficiently absorb gluons with $k \sim T$, and the reverse process plays an indispensable role for thermalization. 
In the next section we shall analyze numerically under what conditions the gluon absorption is important.

A complication of inelastic processes is that the radiated (absorbed) gluon takes a finite amount of time to be fully resolved from (merged into) the parent parton.
This typical time scale is called the gluon formation time $\tau_f$, during which the effects of multiple collisions add up coherently in a destructive manner to suppress the gluon emission spectrum \cite{Wang:1994fx, Baier:1996kr, Zakharov:1996fv}, known as the LPM effect in analogy to QED \cite{PhysRev.103.1811}.
The formation time for gluons with a thermal mass splitting from the heavy quark is \cite{Cao:2013ita},
\begin{eqnarray}
\tau_f &=& \frac{2x(1-x)E}{k_\perp^2 + x^2M^2 + (1-x)m_D^2/2}, x = \frac{k+k_z}{E+p_z}.
\end{eqnarray}
For certain phase space regions of the radiated gluon, this time scale can be comparable to or even much larger than the mean free path $\lambda$, making this a non-local task in a Monte Carlo approach.
This creates a paradox in the Boltzmann equation formulation where all scatterings are point-like (compared to $\lambda$) in space-time.
A possible solution to this paradox has been proposed by \cite{ColemanSmith:2012vr}: it suggests treating this long-lived system of heavy quark plus primitive gluon as a continuum specie of "particles" that can propagate within its life time and scatter with medium partons.
However in this work, we still treat the radiation in a point-like interaction manner for simplicity and mimic this LPM effect by restricting the phase space integral of the emission / absorption gluon with a coherence factor,
\begin{eqnarray}\label{eq:LPM}
dk^3/2k \rightarrow 2\left(1 - \cos\left((t-t_0)/\tau_f\right) \right)dk^3/2k,
\end{eqnarray}
where $t-t_0$ is the time elapsed from the last emission / absorption.
This factor is determined by requiring that the differential radiation rate reduces to the higher twist formula used by \cite{Cao:2013ita} in the limit that gluon is soft ($x\ll 1$) and its transverse momentum is much larger than the medium momentum transfer ($k_\perp^2 \gg q_\perp^2$).
With this prescription, the rate of emitting a gluon with $\tau_f \gg t-t_0$ is suppressed as required at the expense of the collision rate becoming history ($t-t_0$) dependent.
It is not trivial to ascertain whether the Boltzmann equation with a history-dependent rate should thermalize as $t\rightarrow \infty$, so we test and confirm this in the next section.

\subsection{Diffusion component}
The Fokker-Planck part of the transport equation can be solved by propagating heavy quarks using Langevin equations in between subsequent pQCD scatterings.
The Langevin equations in the pre-point Ito scheme are \cite{Rapp:2009my},
\begin{eqnarray}
\Delta \vec{x}_i &=& \frac{\vec{p}_i}{E} \Delta t	\\
\Delta \vec{p}_i &=& -\eta_D \vec{p}_i \Delta t + \Delta t \vec{\xi}(t)
\end{eqnarray}
The first equation is the spatial transport.
In the second equation, the momenta of heavy quarks are changed by a drag term with coefficient $\eta_D$ and a thermal random force $\vec{\xi}$. 
The random force has zero mean and the covariance structure:
\begin{eqnarray}
\langle \xi_i \xi_j \rangle = B_{ij} = \frac{\Kpara}{\Delta t} \frac{p_i p_j}{p^2} + \frac{\Kperp}{\Delta t} \left(\delta_{ij} - \frac{p_i p_j}{p^2}\right)
\end{eqnarray}
In this study, we assume an isotropic non-perturbative diffusion $\Kpara=\Kperp=\kappa$.
The drag coefficient $\eta_D$ and the momentum diffusion coefficient $\kappa$ need to satisfy the Einstein relation in the pre-point Ito scheme to guarantee the approach of the correct thermal equilibrium \cite{Rapp:2009my},
\begin{eqnarray}
\eta_D &=& \frac{\kappa}{2TE} - \frac{d\kappa}{dp^2}
\end{eqnarray}
We choose to parametrize the momentum diffusion constant $\kappa$ and the drag is determined as above.
If temperature is the only scale in the problem, we expect $\kappa$ to scale as $T^3$.
It is also natural to expect non-perturbative contribution may be large at low energy and low temperature, so we arrive at the following simple ansatz,
\begin{eqnarray}
\frac{\kappa}{T^3} = \kappa_D\left(x_D + (1-x_D)\frac{\textrm{ GeV}^2}{ET}\right).
\end{eqnarray}
Here, $\kappa_D$ is the strength of diffusion at $ET = 1\textrm{ GeV}^2$, and $x_D$ interpolates between the two types of energy-temperature dependence.

\section{Tests in a static medium}\label{section:test}
\begin{figure}
\includegraphics[width=\columnwidth]{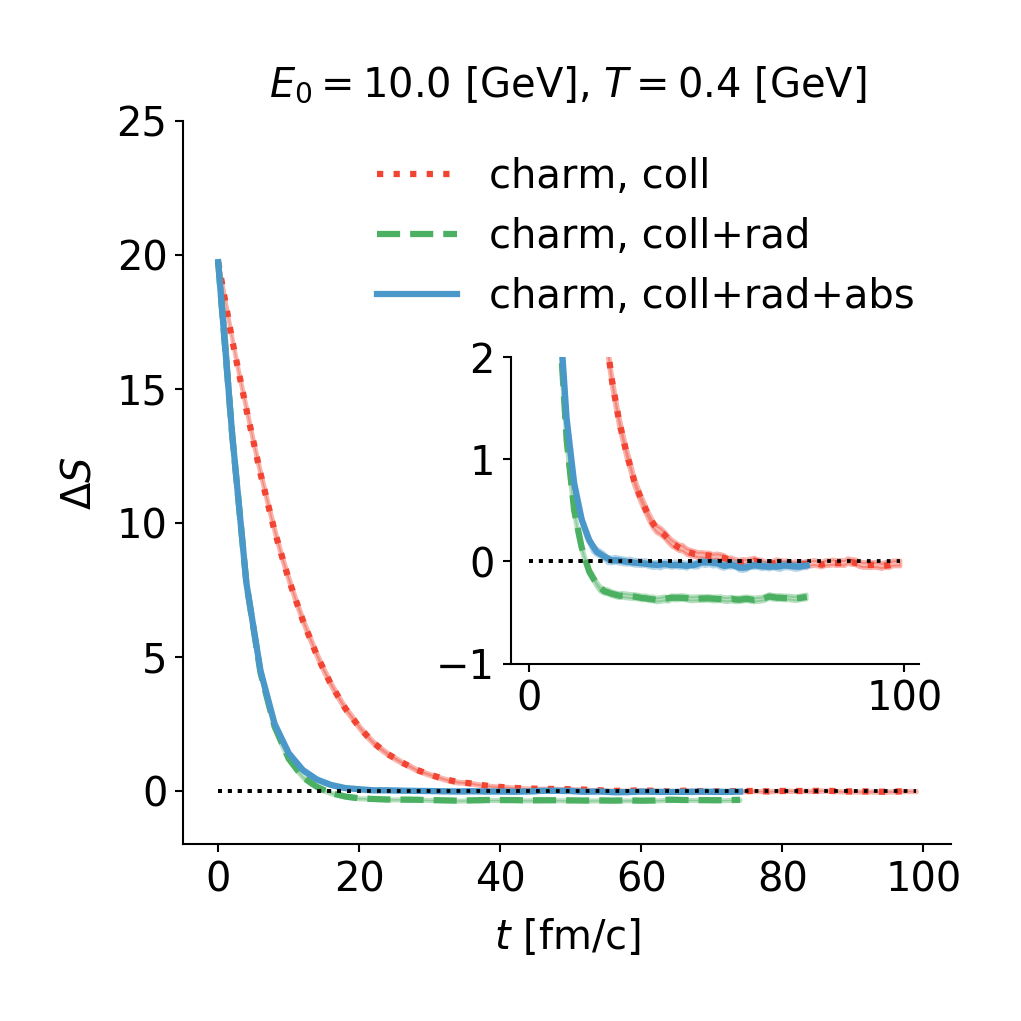}
\caption{The approach to thermalization of the linear Boltzmann equation with elastic processes only (red dot), elastic and radiation processes (green dashed), and elastic with both radiation and absorption processes (blue solid). The static medium has a temperature $T = 0.4$ GeV. $10^4$ heavy quarks are initialized with $E = 10$ GeV at $t = 0$.}\label{plots:thermalization}
\end{figure}

Before coupling the heavy quark transport model to a realistic medium, we study the model in a static medium, i.e., the medium is at rest with a fixed temperature.
All the calculations in this section use $\mu =1$ and $\kappa = 0$.

A first test is to check the implementation of detailed balance to see whether the system reaches the proper thermal equilibrium.
To quantify the approach of an ensemble of $N$ heavy quarks to a thermal distribution in a medium with temperature $T_0$, we define the following indicator $\Delta S$,
\begin{eqnarray}
\Delta S = \frac{1}{N}\sum_{i=1}^{N} \ln f_0(E_i) - \frac{\int f_0(p)\ln f_0(E) dp^3}{\int f_0(p) dp^3}
\end{eqnarray}
$f_0 \propto \exp(-E/T_0)$ is the Boltzmann-J\"uttner distribution function. 
The first term takes the heavy quarks ensemble average of $\ln f_0(E)$ and the second term is proportional to the entropy at $T_0$,
This difference $\Delta S$ defines a ``distance" of the heavy quark ensemble to the thermal distribution, and it vanishes when the ensemble thermalizes.
If the ensemble distribution function $f$ is not far from equilibrium and can be characterized by an effective temperature $T_{\textrm{eff}}$ so that $f(E)\sim \exp(-E/T_{\textrm{eff}})$, then this ``distance" measures,
\begin{eqnarray}
\nonumber
\Delta S &\sim& \frac{1}{T_0}\int  e^{-E/T_{\textrm{eff}}} E dp^3 - \frac{1}{T_0}\int e^{-E/T_0} E dp^3 \\
&=& \frac{T_\textrm{eff}-T_0}{T_0},
\end{eqnarray}
which is the fractional deviation of the effective temperature from the temperature of the thermal bath.
Figure \ref{plots:thermalization} shows the time-evolution of $\Delta S$ of $10^4$ charm quarks inside a thermal bath of $T_0=0.4$ GeV with initial energy $E_0 = 10$ GeV.
With elastic process only, the system thermalizes after about $50$ fm/$c$.
If we now include radiative processes, the system reaches equilibrium faster, but it is the wrong equilibrium.
The effective temperature is lower than the temperature of the thermal bath $T_0$.
This is the consequence of breaking detailed balance without the reverse process of gluon absorption.
Finally, we show the case with both radiation and absorption turned on -- here the correct equilibrium is reached after $t\sim 20$ fm/$c$.
The absorption processes only make a notable difference when the system is not far from equilibrium ($\Delta S < 1$), which is expected from our previous discussion.
\begin{figure}
\includegraphics[width=\columnwidth]{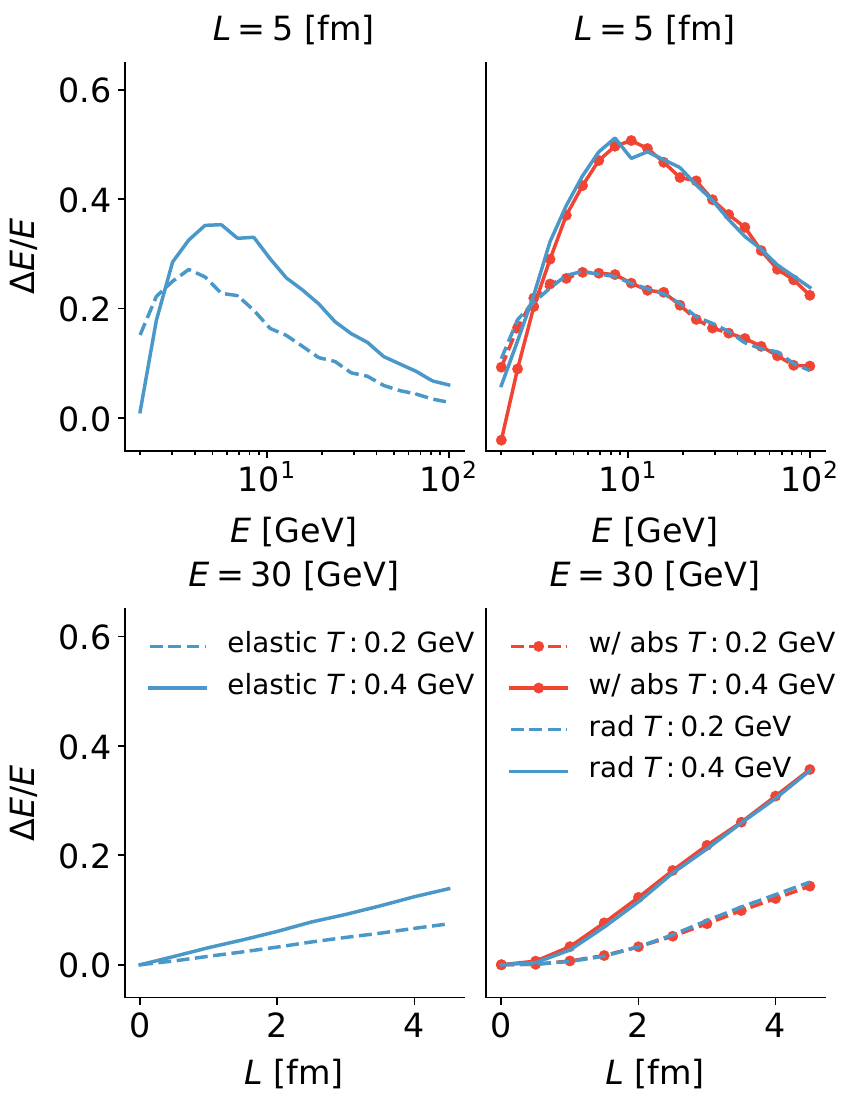}
\caption{Top row: energy loss fraction as function of energy for elastic processes (left) and inelastic processes (right). Bottom row: energy loss fraction as function of path length for elastic processes (left) and inelastic processes (right).}\label{plots:dEE}
\end{figure}
\begin{figure}
\includegraphics[width=\columnwidth]{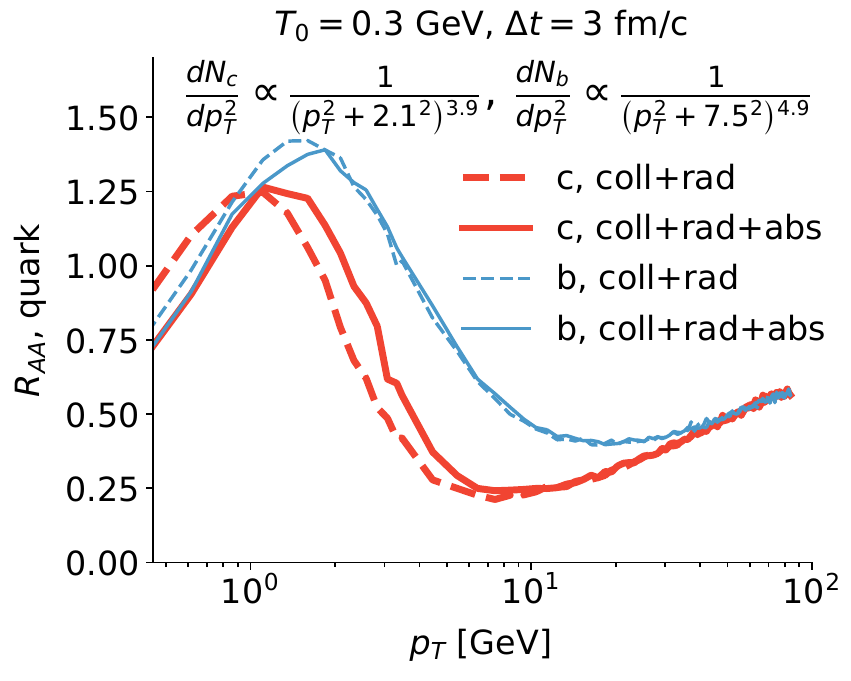}
\caption{$R_{AA}$ in a static medium for charm and bottom quarks. The medium size  is $3$ fm and has a temperature $T_0 = 0.3$ GeV. The initial state spectrum of charm and bottom quarks for this calculation is a simple power law form with fit parameters from \cite{Cao:2012jt}.}\label{plots:BoxRaa}
\end{figure}

Next we study heavy quark energy loss in a static medium.
Na\"ively, energy loss per unit time can be calculated by inserting $\Delta E$ into the integration of the rate equation \ref{eq:rate}. 
This is straightforward for elastic processes, but since the rates of inelastic processes depend on the interaction history, a meaningful energy loss can only be calculated by performing an actual Monte Carlo simulation.
And as we will see, this interaction history dependence causes a non-trivial path length (medium size $L$) dependent energy loss.
In the first row of Figure \ref{plots:dEE}, we show the energy loss fraction $\Delta E/E$ for elastic processes (left) and inelastic processes (right) as function of $E$ for a path length of $5$~fm at temperatures of $T=0.2$ and $0.4$  GeV.
The elastic energy loss fraction is large at intermediate energy and decreases towards small and large energies.
At sufficiently low energy, the heavy quark starts to gain energy from the medium on average which manifests as $\Delta E/E < 0$.
For the case of inelastic energy loss, we study the effect of the gluon absorption process by comparing $\Delta E/E$ with only radiation processes (lines) to $\Delta E/E$ with both gluon radiation and absorption (lines with symbols).
As expected, we find that the gluon absorption process only affects energy loss significantly for small values of $E/T$.
At sufficiently low energy, the gluon absorption process allow the heavy quark to gain energy from the medium through the inelastic channel which is key to thermalization.
In the second row of Figure \ref{plots:dEE}, we show the path length dependence of the two energy loss mechanisms.
Here, we plot the energy loss fraction per unit length.
The key observation is that the elastic energy loss increases linearly with path length but the inelastic energy loss increases non-linearly for small path length and then transits to a linear increase at large path length. 
The non-linear $L-$dependence is a characteristic behavior of the coherence effect in a finite length medium. 
In our effective LPM implementation, this arises because gluon radiation with $\tau_f \sim k/T^2 \gg L$ is suppressed.
Therefore for a thin medium, the phase space for gluon radiation is restricted $k < LT^2$, which is also the typical amount of energy loss per radiation.
Multiplying $k \sim LT^2$ by the number of collisions $N \propto LT$, the inelastic energy loss scales as $\Delta E \sim L^2T^3$.
For a thick medium, a heavy quark could have multiple radiations with $N \propto L/\tau_f$ and each radiation carries off a typical amount of energy. 
In this region, the inelastic energy loss rises linearly with $L$.
What we see in the simulation is a behavior that interpolates between these two qualitative behaviors.

Finally, we calculate the nuclear modification factor $R_{AA}$ of charm and bottom quarks in a static medium $T=0.4$ GeV after evolving for $3$ fm/$c$.
The initial spectra of the charm and bottom quarks are parametrized in a simple power law form \cite{Cao:2012jt}.
This simplified setup is intended for comparison with other models with controlled settings.
In Figure \ref{plots:BoxRaa} we show both $R_{AA}$ of charm and bottom with or without the gluon absorption process. 
Again, we see that the absorption process only affects observables for relatively low momenta $p_T < 10$ GeV.
The mass plays an important role in the intermediate $p_T$ region, where a clear separation between charm and bottom $R_{AA}$ is visible.
The mass effect looses importance at high energy where $p_T$ is the only relevant scale or at low $p_T$ when $M \gg p_T$ and $T$. 

\section{Model calibration using Bayesian analysis}\label{section:calibration}
Finally, we couple our transport model to a state-of-the-art 2+1D event-by-event viscous hydrodynamical medium evolution and extract the model parameters from a Bayesian model-to-data comparison.
The medium evolution model consists of multiple stages,
\begin{itemize}
\item[1.] The \trento\ model generates event-by-event initial conditions at time $\tau = 0^+$ \cite{Moreland:2014oya}. 
\item[2.] A collision-less Boltzmann equation (free streaming) models the pre-equilibrium stage prior to the start of the hydrodynamic evolution at $\tau_{fs}$ \cite{Liu:2015nwa}.
\item[3.] 2+1D event-by-event relativistic viscous hydrodynamics evolves the QGP with an up-to-date lattice equation of state (EoS) \cite{Shen:2014vra, Bazavov:2014pvz}.
\item[4.] Finally, hadrons sampled from hydrodynamic energy momentum tensors rescatter and decay in the hadronic phase utilizing the Ultra-Relativistic Quantum Molecular Dynamics (UrQMD) model \cite{Bass:1998ca, Bleicher:1999xi}.
\end{itemize}
The parameters of this particular bulk medium evolution have already been calibrated to reproduce a vast array of bulk observables at LHC energies \cite{Bernhard:2018hnz}, providing a description of the bulk evolution of the QGP with unprecedented precision.
On the heavy quark transport side, we initialize heavy quark ensembles with momenta sampled from a FONLL calculation using two different sets of nuclear parton distribution functions (PDFs) \cite{Cacciari:1998it,Kovarik:2015cma,Eskola:2016oht}. 
The nuclear PDFs comes with a large uncertainty in the shadowing region which is relevant at the LHC energies.
It is hard to systematically include this uncertainty in our study; instead, we choose to use the center values of two different sets of nuclear PDFs, namely the {\tt EPPS} set and the {\tt nCTEQ16np} set and perform calibrations using both to demonstrate the sensitivity of the parameter extraction on the nuclear PDF uncertainty.
The position of the hard production vertices at $\tau = 0^+$ are sampled from the \trento\ binary collision density to correlate with hot-spots of  underlying event. 
During the pre-equilibrium stage, the heavy quarks should already start to interact with medium.
However, the system at this stage is still far off both kinetic and chemical equilibrium.
To get a handle on the effect of pre-equilibrium energy loss, we choose to define the medium flow velocities and energy density from the pre-equilibrium energy-momentum tensor by Landau matching and convert the energy density to an effective temperature using a three-flavor conformal QCD EoS. 
Heavy quarks are allowed to loose energy from a tunable energy-loss starting time $0.1< \tau_0 < 1.0 \textrm{ fm/$c$} $. 
With a small $\tau_0$, this correspond to a fast generation of color degrees of freedom in the medium that can collide with heavy quarks at very early times, and with a large $\tau_0$ the pre-equilibrium effects are gradually turned off.
This is of course a rather crude setup and in the future we plan on utilizing more sophisticated models based on kinetic theory to treat pre-equilibrium stage energy loss \cite{Srivastava:2017bcm}.
During the hydrodynamic expansion, the evolution of the flow velocities and temperature are provided by the 2+1D viscous hydrodynamics with boost-invariance in the beam direction.
The heavy quarks subsequently hadronize using a sudden-approximation at $T = 0.154$ GeV via fragmentation and recombination mechanisms \cite{Cao:2013ita}. 
B mesons cease to interact at this point in our model, but D mesons are included in the UrQMD afterburner with $\pi$-D and $\rho$-D cross-sections \cite{Lin:2000jp}.

The parameters of our model in the heavy-flavor sector are:
\begin{itemize}
\item[1.] $\tau_0$ the time at which heavy quark energy loss starts, varying between $0.1$ fm/$c$ to $1.0$ fm/$c$,
\item[2.] $\mu$, the medium energy scale ($\mu\pi T$) that appears in the running coupling constant of the scattering component, varying from $1/3$ to $4$,
\item[3.] $\kappa_D$, the strength of momentum diffusion at $ET = 1 \textrm{GeV}^2$, ranging from $0$ to $8$, and
\item[4.] $x_D$, the fraction of the momentum diffusion that is energy independent, ranging from $0$ to $1$.
\end{itemize}
In addition to these continuous parameters, the choice of different nuclear parton distribution functions acts as a discrete variable.

We now briefly introduce the Bayesian techniques and key terminologies to be used later. These techniques have been described in great detail in a series of publications regarding their application to the extraction of bulk QGP properties and initial conditions of heavy-ion collisions \cite{Bernhard:2015hxa,Bernhard:2016tnd} as well as in the heavy quark sector to the extraction of the heavy quark diffusion coefficient within the framework of an improved Langevin transport model \cite{Xu:2017obm}.
The application of these techniques to relativistic heavy-ion collisions in general has been part of the 
thesis work by J. Bernhard \cite{Bernhard:2018hnz}.

Given a model whose prediction ${\bf y}$ depends on a vector of input parameters ${\bf p}$ and  experimental data ${\bf y}_\textrm{exp}$, 
the probability distribution of the {\it true} model parameters ${\bf p^*}$ is given by Bayes' theorem, 
\begin{eqnarray}\label{eq:Bayes}
\textrm{Posterior}({\bf p^*}|{\bf y}_\textrm{exp}, M) &\propto& \textrm{Likelihood}({\bf y}_\textrm{exp}|{\bf p^*}, M) \nonumber \\ &\times& \textrm{Prior}({\bf p^*}).
\end{eqnarray}
The posterior probability distribution of the ${\bf p^*}$ given a certain model $M$ and data, equals the probability $L$ of observing the data given the model and parameters ${\bf p^*}$, called likelihood function, times a prior belief on the distribution of ${\bf p^*}$.
The likelihood function is often defined in a Gaussian form in terms of the difference between model calculation and experimental data and a covariance matrix $\Sigma$ that encodes experimental and theoretical uncertainties,
\begin{eqnarray}\label{eq:likelihood}
\ln(L) &=& -\frac{1}{2}({\bf y}-{\bf y}_{\textrm{exp}})^T\Sigma^{-1} ({\bf y}-{\bf y}_{\textrm{exp}})\nonumber\\ 
		&-&\frac{d}{2}\ln(2\pi)-\frac{1}{2}\ln|\Sigma|.
\end{eqnarray}
The construction of $\Sigma$ is described in Appendix \ref{appendix:sigma}.
Once we have the ability to evaluate model output given arbitrary parameters within a reasonable range, the information on the parameters constrained by data follows from Equations (\ref{eq:Bayes}) and (\ref{eq:likelihood}).
This high-dimensional posterior probability distribution function can be sampled using a Markov-chain Monte Carlo (MCMC) procedure.
The main challenge for applying this method directly to event-by-event heavy-ion collision models resides in the computational effort required for the model calculations.  $\mathcal{O}(10^4)$ minimum-biased events are needed to get statistical uncertainties of the calculation under control.
Since it is impractical to evaluate the model  at arbitrary points in parameter space during the MCMC sampling, alternative methods for rapid model evaluations have to be found.
The solution is to use an advanced sampling technique by only evaluating the full model at $\mathcal{O}(100)$ design parameter sets (design points) and subsequently interpolating the model to generate output at arbitrary points in parameter space using  Gaussian process emulators that have been trained on the full model calculation at the design points \cite{Rasmussen:2006gp}.
\begin{center}
\begin{table}[h]
\caption{ALICE dataset}\label{table:ALICE-obs} 
\begin{tabularx}{\columnwidth}{XXX}
\hline 
 Observables & Centrality & Reference\\ 
\hline 
$D$-meson $v_2$ & 30-50\% & \cite{Acharya:2017qps}\\ 
\hline 
Event-engineered $D$-meson $v_2$ & 30-50\% & \cite{Grosa:2017zcz}\\ 
\hline 
$D$-meson $R_{AA}$ & 0-10, 30-50, 50-80\% & \cite{Acharya:2018hre}\\
\hline 
\end{tabularx}
\end{table}
\begin{table}[h]
\caption{CMS dataset}\label{table:CMS-obs} 
\begin{tabularx}{\columnwidth}{XXX}
\hline 
Observables & Centrality & Reference\\ 
\hline 
D${}^0$-meson $v_2$ & 0-10, 10-30, 30-50\% & \cite{Sirunyan:2017plt}\\ 
\hline 
D${}^0$-meson $R_{AA}$ & 0-10\%, 0-100\% & \cite{Sirunyan:2017xss}\\ 
\hline 
B${}^{\pm}$-meson $R_{AA}$ & 0-100\% & \cite{Sirunyan:2017oug}\\ 
\hline 
\end{tabularx}
\end{table}
\end{center}

In this work, we sampled 80 design points in a four dimensional parameter space $(\tau_0, \mu, \kappa_D, x_D)$.
For each parameter set, we run 4000 minimum bias events.
Each event propagates an ensemble of $4\times 10^4$ charm quarks and $10^4$ bottom quarks.
The centrality is defined by the mid-rapidity charged particle multiplicity and the same kinematic cuts as are used by the experiments are applied to the calculation of heavy-flavor observables.
All observables are measured at 5.02 TeV in Pb+Pb, as listed in Table \ref{table:ALICE-obs} and \ref{table:CMS-obs}. Most of the data we utilize are for $D$-mesons: the
 $p_T$ dependent $D$-meson nuclear modification factor $R_{AA}$ and $p_T$ dependent second-order azimuthal anisotropy $v_2$ at various centralities \cite{Sirunyan:2017plt, Sirunyan:2017xss, Acharya:2017qps,Grosa:2017zcz}.
We also compare to the event-shape-engineered $D$-meson $v_2$ measured by the  ALICE collaboration \cite{Grosa:2017zcz}.
The idea of the event-shape engineering is to subdivide events at a certain centrality according to the magnitude of the charged particle $q$-vector, in this case,
\begin{eqnarray}
|q_2|^2 = \frac{\left(\sum_{i=1}^{M} \cos(2\phi) \right)^2+ \left(\sum_{i=1}^{M} \sin(2\phi) \right)^2}{M} \, .
\end{eqnarray}
The $D$-meson $v_2$ is measured for those events with $20\%$ highest $q_2$ and events with $60\%$ lowest $q_2$.
It is found that $D$-meson flow is strongly correlated with this measurement of bulk collectivity.
This event-shape-engineering procedure necessitates a full event-by-event study and may be sensitive to the interplay between heavy quark energy loss and initial condition fluctuations, so we include this observable into the set of observables on which we calibrate the model.
Finally in order to require the calibrated model to predict the desired mass-dependence, we include recent CMS measurements of $B^{\pm}$-meson $R_{AA}$, although the data have a large uncertainty which suppresses its importance in the likelihood function.
\begin{figure*}
\includegraphics[width=.49\textwidth]{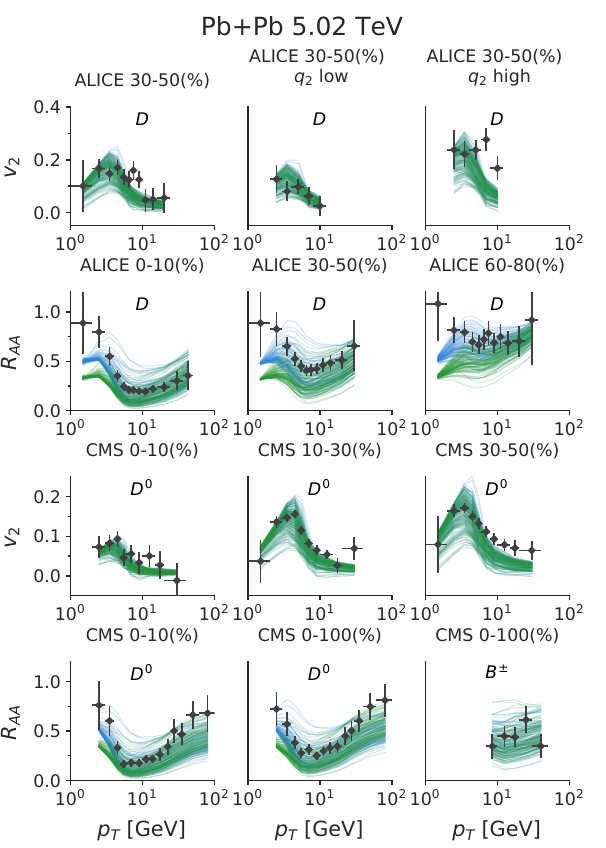}
\includegraphics[width=.49\textwidth]{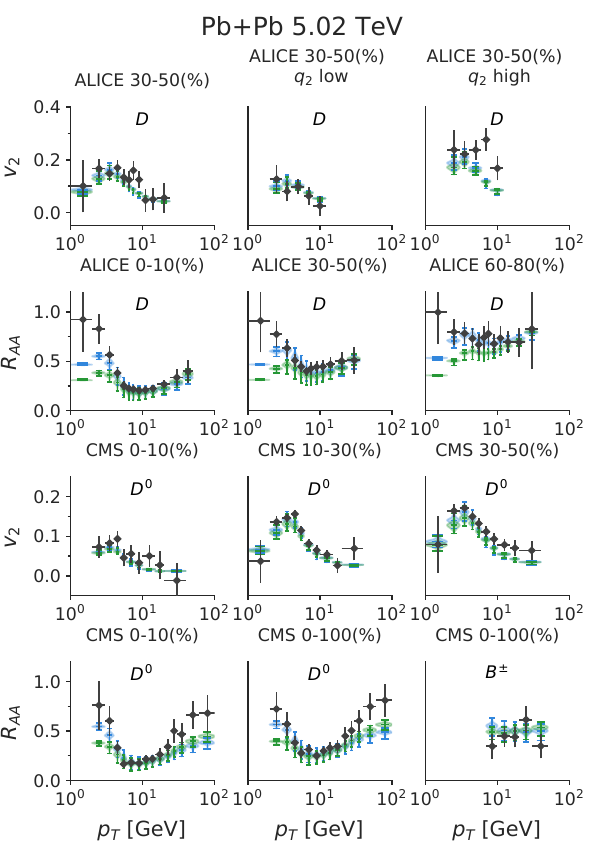}
\caption{Left: the prior, i.e. the full range of calculations in parameter space. Right: the posterior, i.e. observables sampled from model emulators after calibration. In both figures, blue (green) lines are calculations with {\tt EPPS} ({\tt nCTEQ15np}) nuclear PDF.}\label{plots:deisgn_posterior_obs}
\end{figure*}

On the left of Figure \ref{plots:deisgn_posterior_obs}, we show the prior, i.e. the full range of our calculations in parameter space for each of the listed observables. 
We use different colors to distinguish calculations using {\tt EPPS} (blue) and
{\tt nCTEQnp} (green) nuclear PDF.
The calculated values of $R_{AA}$ at high transverse momenta and $v_2$ at low transverse momenta have a large spread, sufficiently wide to cover the experimental data.
We notice that the model always underestimates very low-$p_T$ $R_{AA}$ points and 30-50\% high-$p_T$ $v_2$ from CMS.
This could be a limitation of our model, such as the need of a more sophisticated implementation of the LPM effect or the need for a more accurate calculation of initial low-$p_T$ charm quark production in both $p$-$p$ and $A$-$A$ collisions. 
Plots on the right of Figure \ref{plots:deisgn_posterior_obs} show the posterior distribution of the observables from model emulators, i.e. interpolated model predictions after calibration.
The calibrated model displays a very good overall agreement with all the observables except for the cases pointed out above.
The use of different nuclear PDFs has a  negligible effect on azimuthal anisotropy observables, but does affect the $R_{AA}$ at small and large $p_T$.
Calculations with the {\tt EPPS} nuclear PDF work very well in describing $R_{AA}$ below $p_T = 50$ GeV, while calculations with the {\tt nCTEQnp} do slightly better for CMS $R_{AA}$ data with $p_T>50$ GeV.
The calculated event-engineered flow strongly correlates with the charged particle $|q_2|$ and describes the lowest 60\% $q_2$ bin very well.
For the highest 20\% $q_2$ bin, the model posterior is consistent with measurements below $5$ GeV and underestimates the data at higher $p_T$ bins.

\begin{figure}
\includegraphics[width=.5\textwidth]{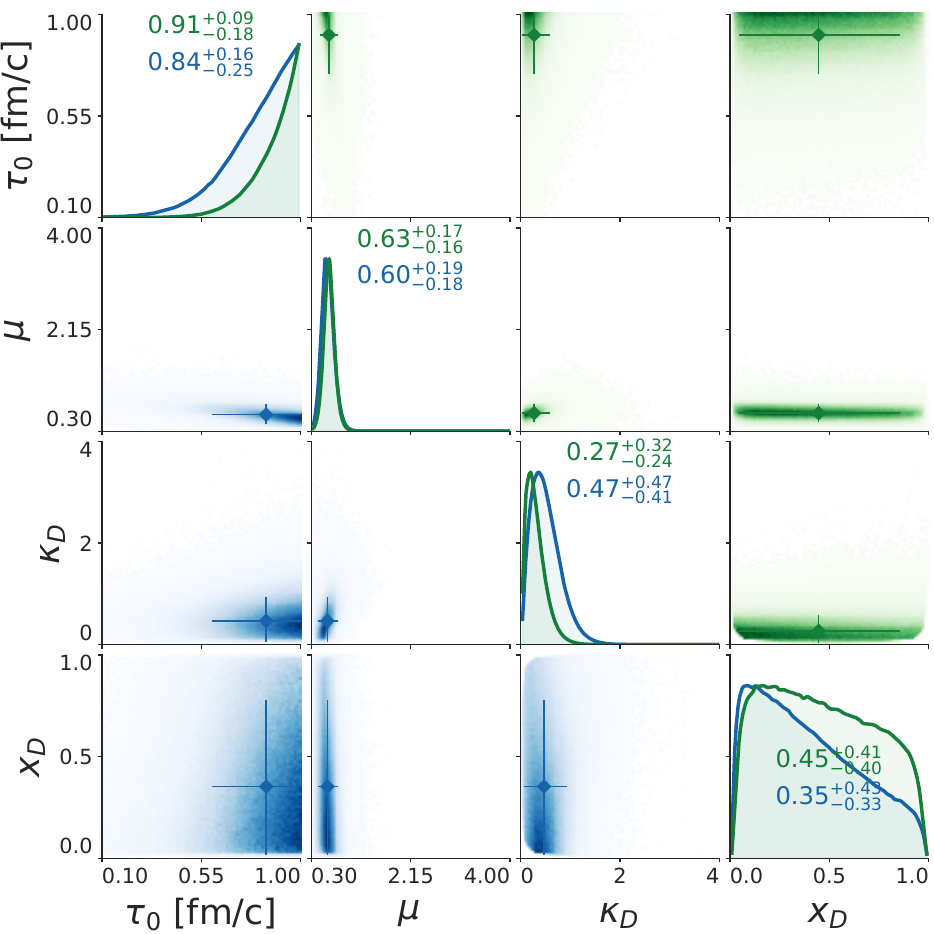}
\caption{Marginalized postrior probability distribution of model parameters. Diagonal plots show the marginalization on a single parameter. Off diagonal plots show the pair correlation between parameters. Blue (Geen) lines and lower (upper) off diagonal plots correspond to the extraction using EPPS (nCTEQ15np) nuclear PDF.}\label{plots:posterior}
\end{figure}
\begin{figure}
\includegraphics[width=.5\textwidth]{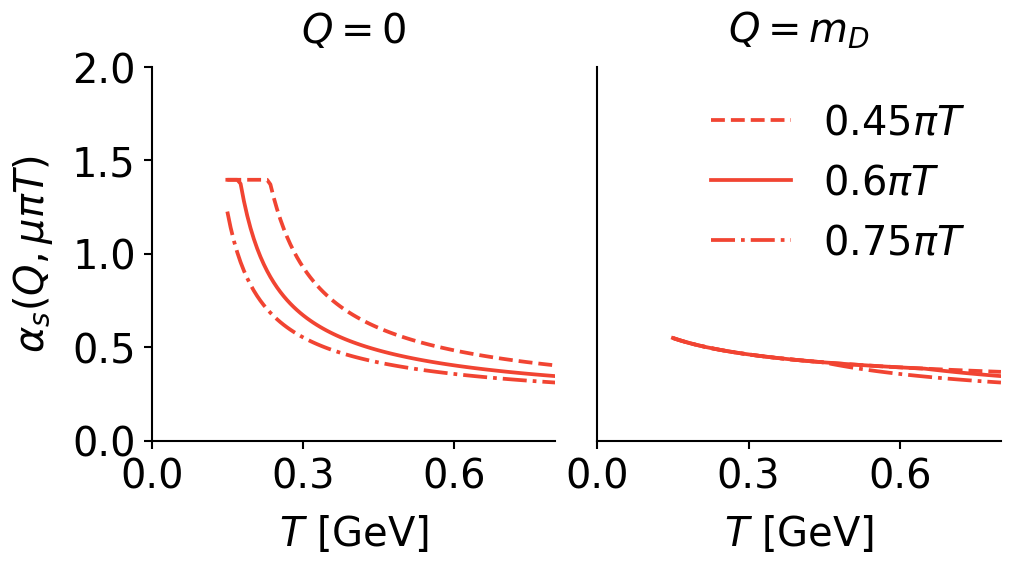}
\caption{Coupling constant with three values of the medium scale parameter $\mu$ taken  from the high likelihood region of the posterior. Left: $\alpha_s$ evaluated at a process scale $Q=0$. Right: $\alpha_s$ evaluated at a process scale $Q=m_D$.}\label{plots:alphas}
\end{figure}

The posterior probability distribution of all parameters is marginalized to single parameter distributions (diagonal) and two-parameter joint distributions (off-diagonal) in Figure \ref{plots:posterior}.
The lower off-diagonal plots and blue lines in the diagonal plots correspond to the calibration using the {\tt EPPS} nuclear PDF, and the upper off-diagonal plots and green lines in the diagonal plots use the {\tt nCTEQ15np} nuclear PDF.
Despite the difference in $R_{AA}$ when different nuclear PDFs are used, the extracted probability densities of parameters are similar.
To describe LHC data, the model prefers a late onset of medium induced energy loss and a medium energy scale roughly around $0.6\pi T$, which implies the largest coupling constant at a given temperature is $\alpha_s \sim \alpha_s(1.8T)$.
A small but finite amount of momentum diffusion at $ET=1\textrm{ GeV}^2$ is preferred for the diffusion component.
The smallness of this number is expected since most of the interaction is already taken account by the pQCD scattering component with a relatively large coupling constant (i.e. a small medium scale).
We find this study to be not sensitive to the energy / temperature dependence of the diffusion component beyond the regular $T^3$ scaling of the momentum diffusion constant.   

The preferred medium scale parameter $\mu \sim 0.6$ is not large which could result in a large $\alpha_s$.
Therefore, we check the range of typical $\alpha_s$ values in the model in order to evaluate the use of perturbative matrix-elements.
Figure \ref{plots:alphas} shows the coupling constant evaluated at two process scales $Q=0$ and $Q=m_D$.
In the case of $Q=0$ (left), the energy scale is cut off by $\mu\pi T$ and this plot show the maximum of model coupling constant at a given temperature.
Setting $Q=m_D$ (right) as a proxy for the typical momentum transfer in the $t-$channel scattering, the coupling constant rises slower as temperature drops.
It is found that in order to describe experimental data, the preferred coupling constant is fairly large, suggesting next-to-leading (NLO) order corrections to the present scattering picture should be prominent. 
Because these large $\alpha_s$ values are encountered in small-momentum-transfer scatterings ($0< Q < m_D$), we will absorb these small-momentum-transfer elastic and inelastic pQCD processes into a radiation-improved Langevin equation in future studies. 
This way, one not only avoids the explicit use of large $\alpha_s$ in pQCD matrix-elements, but also interpolates between the pQCD based scattering model, the radiation-improved Langevin model and pure non-perturbation drag and diffusion model with one or two control parameters, allowing for more systematic model-uncertainty study.

\begin{figure}
\includegraphics[width=\columnwidth]{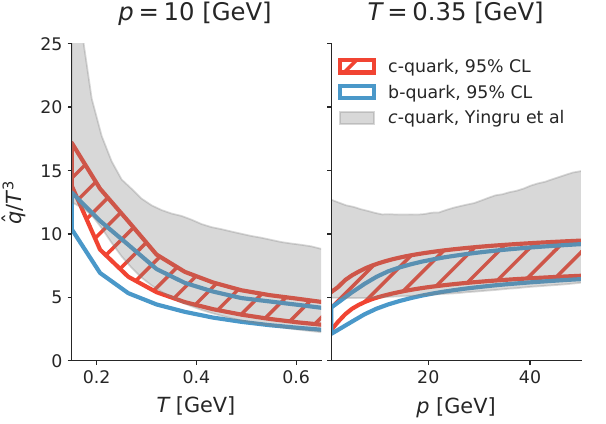}
\caption{Posterior range of the heavy quark transverse momentum broadening parameter $\hat{q}$ from Equation \ref{eq:qhat}. The results include the uncertainty from using different nuclear PDFs. Blue boxed region is for bottom quarks and red slashed region for charm quarks. The shaded region indicates a previous extraction \cite{Xu:2017obm}.}\label{plots:posterior_qhat}
\end{figure}

\begin{figure}
\includegraphics[width=\columnwidth]{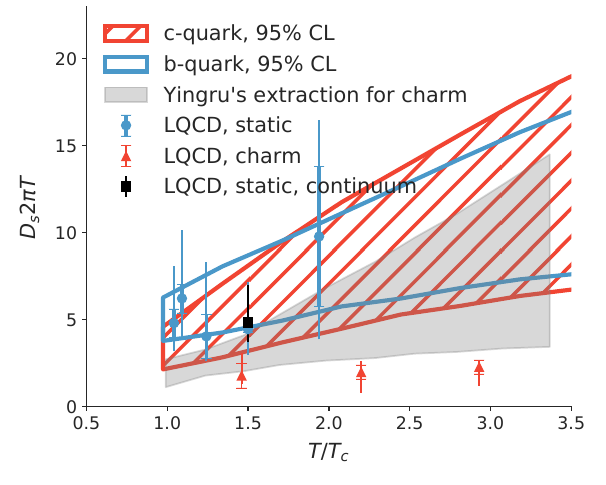}
\caption{Posterior range of the heavy quark spatial diffusion coefficient. The blue boxed region is for bottom quarks and the red slashed region for charm quarks. The shaded region indicates a previous extraction in \cite{Xu:2017obm}. Square and dimond symbols are lattice calculations in the static heavy quark limit \cite{Banerjee:2011ra, Francis:2015daa}; triangular symbols are lattice calculations with physical charm quark mass \cite{Ding:2012sp}.}\label{plots:posterior_Ds}
\end{figure}

Next, we investigate the transport coefficients extracted from the calibrated model.
To define the transport coefficient of a heavy quark,
we combine the contribution from both elastic scatterings and the diffusion component,
\begin{eqnarray}\label{eq:qhat}
\frac{\hat{q}}{T^3} &=& \frac{1}{T^3}\frac{d}{dt}\left\langle p_\perp^2 \right\rangle\\
\nonumber
 &=&  \kappa_D\left(x_D + (1-x_D)\frac{\textrm{GeV}^2}{ET}\right) + \frac{\hat{q}_{\textrm{el}}}{T^3}.
\end{eqnarray}
Where $\hat{q}_{\textrm{el}}$ is obtained by integrating the rate equation with inserting the transverse momentum transfer square.
We shall discuss the inclusion of inelastic process into the calculation of $\hat{q}$ in section \ref{section:conclusion}.
Performing this calculation for many random parameter set samples drawn from the posterior distribution using either nuclear PDF, we determine the posterior distribution of the functional $\hat{q}(E, T)$ constrained by data.
On the left of Figure \ref{plots:posterior_qhat}, we showed the 95\% credible region of $\hat{q}$ as function of temperature, fixing the heavy quark momentum at $10$ GeV.
The right panel of the figure shows $\hat{q}$ as function of momentum at $T=0.35$ GeV.
Our formula includes a mass dependence -- therefore  the charm quark $\hat{q}$ (region enclosed by thick red lines and slashes) is slightly different from the bottom quark $\hat{q}$ (region enclosed by thick blue lines).
The present result is consistent with previous work by Xu et al (shaded region) \cite{Xu:2017obm}, who used an improved Langevin model to extract the charm quark transport properties at the LHC,  but hits the lower half of the 95\% credible region of the previous extraction.
Alternatively, one can present our results in terms of the heavy quark spatial diffusion constant $D_s$ often defined in the limit of $p\ll M$.
It is related to the momentum diffusion parameter by
\begin{eqnarray}
2\pi T D_s = \frac{8\pi T^3}{\hat{q}(p\rightarrow 0, T)} \, .
\end{eqnarray}
In figure \ref{plots:posterior_Ds}, we plot the 95\% credible region of both the charm (region enclosed by red thick lines and slashes) and bottom (region enclosed by blue tick lines) quark spatial diffusion constant as function of $T/T_c$.
The results of this work is systematically higher than the extraction from the former work (shaded region).
There have also been attempts made to calculate the spatial diffusion constant of heavy quarks using lattice QCD: three calculations are available, two are calculated in the static heavy quark limit (blue and black symbols with higher values) \cite{Banerjee:2011ra, Banerjee:2011ra}, one of which performs continuum extrapolation (black square) \cite{Francis:2015daa}; the other result uses a realistic charm quark mass (red triangle symbols with lower values) \cite{Ding:2012sp}.
Our posterior of $D_s$ including the diffusion contribution but with only elastic scattering agrees with the lattice evaluation in the static heavy quark limit.
The effect of including the inelastic scattering in $D_s$ will be discussed in the last section.

To summarize this section, we have performed a Bayesian calibration on the model parameters, yielding
generally good agreement to the data.
Although the use of different nuclear shadowing parametrizations does affect the shape of $R_{AA}$, the extracted parameters are not strongly affected.
The extracted parameters indicate a late onset of medium induced heavy quark energy loss and prefer a small but finite diffusion component.
The transport coefficient $\hat{q}$ and spatial diffusion constant $D_s$ are extracted with $D_s$ being compatible with lattice calculations in the static heavy quark limit. 

\section{Validation and Predictions}\label{section:prediction}
We now apply the calibrated model to predict observables that have already been measured but were excluded in the calibration (validation) and also predict new observables.
In principal, any parameter set sampled according to the posterior probability distribution in the high-likelihood region (95\% credible for example) is equally good to make predictions and the resultant differences represent the systematic uncertainties of the calculation.
For simplicity, we only run a single set of high likelihood parameters listed in Table \ref{table:high-likelihood-parameters} for a large number of events.
\begin{table}
\caption{A high-likelihood parameter set}\label{table:high-likelihood-parameters}
\begin{tabularx}{\columnwidth}{XXXXX}
\hline
Parameters & $\tau_0$ [fm/$c$] & $\mu$ & $\kappa_D$ & $x_D$   \\
\hline
Values & 0.9 & 0.6 & 0.4 & 0.5\\
\hline
\end{tabularx}
\end{table} 
\begin{figure*}
\includegraphics[width=\textwidth]{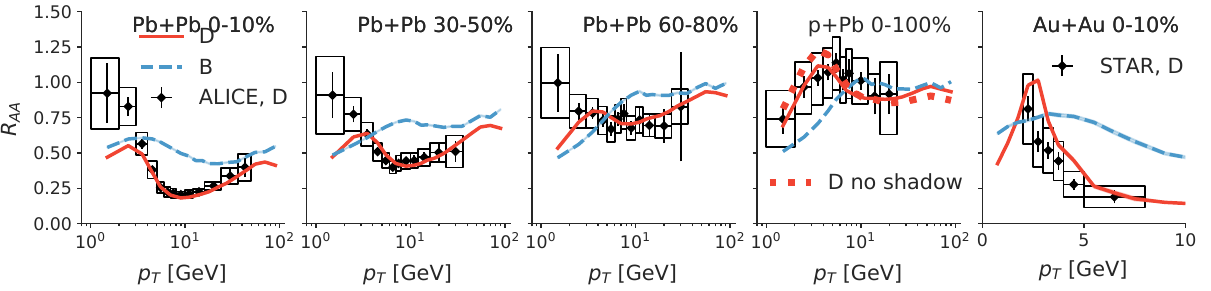}
\caption{Calculation of heavy-flavor $R_{AA}$ using a high likelihood parameter set. Results are compared to ALICE $D$-meson measurements in Pb+Pb and p+Pb \cite{Abelev:2014hha,Abelev:2014hha} and STAR $D$-meson measurements in Au+Au \cite{Xie:2016iwq}. $B$-meson $R_{AA}$ are predicted.}\label{plots:pred:raa}
\end{figure*}
\begin{figure*}
\includegraphics[width=\textwidth]{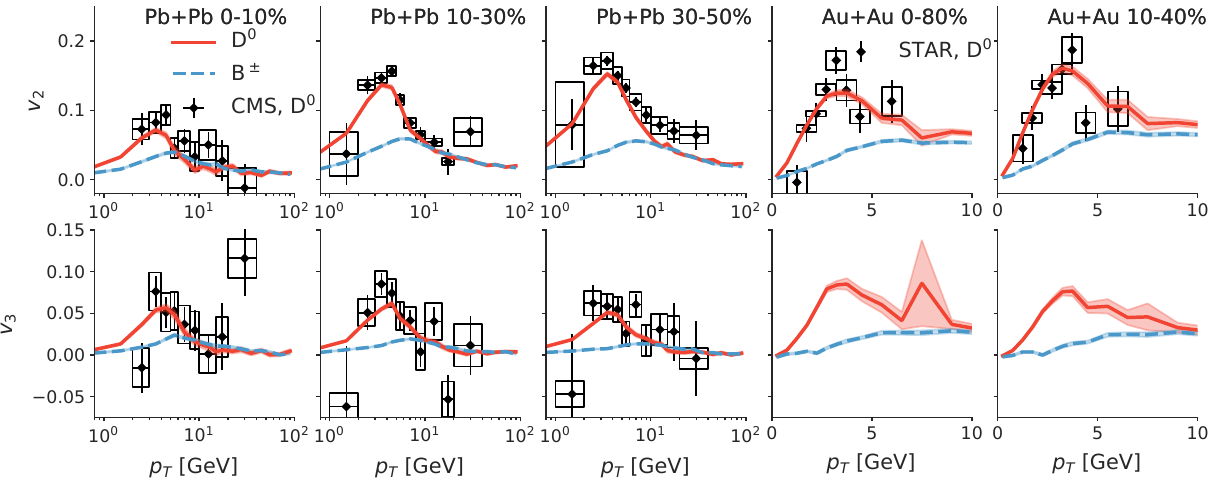}
\caption{Calculation of heavy-flavor flows with a high likelihood parameter set. Results are compared to CMS $D$-meson measurements in Pb+Pb \cite{Sirunyan:2017plt} and STAR $D$-meson measurements in Au+Au \cite{Adamczyk:2017xur}. $D$-meson $v_3$ and $B$-meson $v_2, v_3$ are predictions.}\label{plots:pred:vn}
\end{figure*}
\begin{figure}
\includegraphics[width=0.5\textwidth]{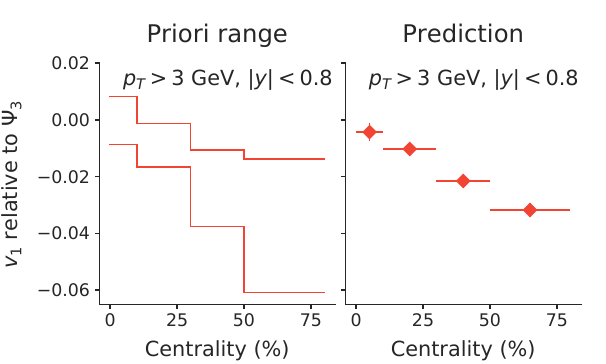}\\
\includegraphics[width=0.5\textwidth]{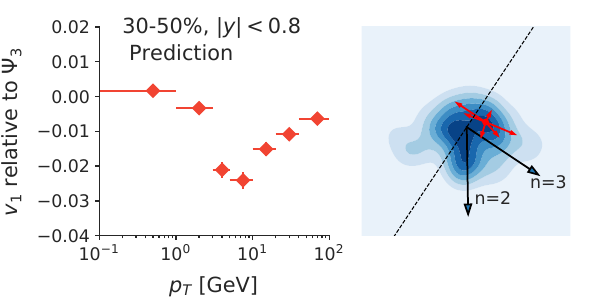}
\caption{Top row: centrality dependence of $D$-meson direct flow with respect to the $n=3$ event plane. Left: the prior range. Right: predictions using a high likelihood parameter set. Bottom left: the transverse momentum dependence of $D$-meson direct flow with respect to the $n=3$ event plane in the 30-50\% centrality bin. Bottom right: a sample TRENTo event. The third order eccentricity would drive a $v_3$. A measurement along the $n=3$ direction would introduce reflection asymmetry in the heavy quark energy loss.}\label{plots:pred:v1}
\end{figure}
We first study the nuclear modification factor in a larger range of $p_T$:
In Figure \ref{plots:pred:raa} the $D$-meson and $B$-meson $R_{AA}$ for 0-10\%, 30-50\%, and 60-80\% centrality are shown and compared to ALICE $D$-meson measurements.
The mass effect clearly separates $B$-meson from $D$-meson $R_{AA}$ in the intermediate $p_T$ range. 
The calculation for $p$+Pb collisions is shown in the fourth plot compared to ALICE measurements \cite{Abelev:2014hha}.
The red-dotted line shows the calculation without nuclear shadowing for $p$+Pb collision.
We find that the calibrated model results are in a very good agreement with the description of the $p+Pb$ minimum bias measurement and that shadowing  is important to understand the low-$p_T$ data.
In the right most plot, we apply the model to Au+Au collisions at RHIC  ($\sqrt{s} = 200$ GeV).
The calculated $D$-meson $R_{AA}$ is slightly higher than the STAR measurement \cite{Xie:2016iwq}, 
yet given that we did not include any RHIC data in our calibration, this level of agreement is satisfactory.

Next, we validate the calibrated model by comparing to CMS measured $D$-meson $v_3$ and make a prediction for $B$-meson $v_n$ in the first three columns of Figure \ref{plots:pred:vn}.
In the transport model, non-zero heavy-flavor $v_3$ is caused by heavy quarks losing energy to a medium that contains a third order eccentricity from initial condition fluctuations.
The $B$-meson $v_2$ and $v_3$ is predicted to be similar to $D$-meson flow for $p_T > 10$ GeV, below which $B$-meson flow is significantly smaller than $D$-meson flow.
Compared to CMS data, the calibrated model reproduces the transverse momentum and centrality dependence of $D$-meson $v_3$ very well.
In the last two columns of Figure \ref{plots:pred:vn}, we again apply the model to the RHIC data and observe a good agreement with STAR measured $v_2$ of $D^0$-mesons \cite{Adamczyk:2017xur}.

Finally, we investigate $D$-meson direct flow $v_1$.
$D$-meson $v_1$ is tiny if one measures it with respect to the reaction plane due to the reflection symmetry on averaging over multiple events.
However, correlating heavy quark $v_1$ with charged particle $v_3$ results in a non-zero signal even at mid-rapidity.
This directed flow with respect to the $n=3$ event plane is calculated in the scalar product approach
\begin{eqnarray}
v_1 &=& \left\langle \frac{\Re\{q_3 Q_1^*\}}{mM} \right\rangle\Bigm/\sqrt{\left\langle \frac{|q_3|^2-m}{m(m-1)} \right\rangle}, \nonumber\\
Q_1 &=& \sum_{j=1}^{M}e^{i\phi_j} \textrm{, for heavy mesons},\nonumber\\
q_3 &=& \sum_{j=1}^{m}e^{i3\phi_j} \textrm{, for charged particles}. 
\end{eqnarray}
The resulting $v_1$ as function of centrality is shown in Figure \ref{plots:pred:v1}.
The upper left plot shows the prior range of this quantity as function of centrality using events from the 80 design points parameter set calculations.
The upper right dots are our prediction using the selected high likelihood parameter set.
The calculated $v_1$ is clearly finite and negative in our calculation and we expect it to reach as far as $-3\%$ in the peripheral centrality bin.
The transverse momentum dependence of $v_1$ is shown in the lower left plot, the magnitude of the signal grows with $p_T$ until it reaches the minimum at $p_T \sim 8$ GeV; then the magnitude slowly drop to zero at large $p_T$.
We understand this finite $v_1$ in the following way
(see also the sketch in the lower right of Figure \ref{plots:pred:v1}):
when $q_3$ is finite, the direction of $q_3$ defines a plane to which the medium evolution is reflection asymmetric and this asymmetry causes heavy quarks to loose energy differently depending on the direction of motion being along or against the direction of $q_3$.
Since $q_3$ originates from triangular initial state fluctuations, a finite signal would be another indication of heavy quark energy loss coupling to initial condition fluctuations and bulk collectivity.

\section{Discussion and Conclusion}\label{section:conclusion}
\begin{figure}
\includegraphics[width=0.5\textwidth]{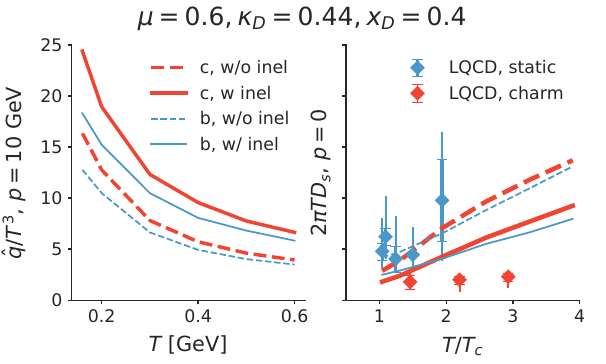}
\caption{Comparison of the heavy quark transport coefficient (left: $\hat{q}$, right $D_s$) with (solid lines) and without (dashed lines) the inclusion of the inelastic contribution. Thin blue lines denote the bottom quark and thick red lines denote the charm quark region. Blue symbols are lattice calculations in static heavy quark limit \cite{Banerjee:2011ra} and red symbols are lattice calculations with physical charm quark mass \cite{Ding:2012sp}.}\label{plots:transport_full}
\end{figure}
Before summarizing this work, we want to return to the definition of $\hat{q}$ in our model.
In Equation (\ref{eq:qhat}), we  only included the elastic scattering processes in the second term and did not include inelastic scattering contributions to momentum broadening.
This is due to having an order-by-order definition of $\hat{q}$ in pQCD.
In fact, the inelastic scattering processes also contain a diffusion-like part, but this contribution is shown to be one order higher in $\alpha_s$ \cite{Ghiglieri:2015ala}, though it may not be numerically small in a realistic scenario compared to leading order.
Since it was our goal to determine a leading order transport coefficient we justify the use of Equation (\ref{eq:qhat}) and neglect any contributions that can be attributed to higher order corrections. 
This choice is also conceptually cleaner for a comparison with other pQCD based calculations.
For lattice calculations the results do not rely on an expansion in $\alpha_s$.
In that case, in order to make a reasonable comparison with lattice transport coefficients, we should calculate $\hat{q}$ from the calibrated model including the inelastic scattering processes.
Unfortunately, currently there is no lattice calculation of $\hat{q}$ available at finite heavy quark momentum.  
$D_s$ on Lattice does exist, but the appropriate momenta ($p \ll M$) are too small for our calculation with Gunion-Bertsch matrix-elements to be valid.
Even so, we would like to show the $\hat{q}$ and $D_s$ with a set of high likelihood parameters that include the inelastic contribution.
Just as the energy loss shown in Figure \ref{plots:dEE}, the transverse momentum broadening per unit time in the presence of inelastic collision processes is not constant for a thin medium.
Therefore, we set up a Monte Carlo simulation for heavy quarks at fixed energy  and extract $\hat{q}$ only after the finite path length effect fades away. 
Figure \ref{plots:transport_full} compares the $\hat{q}$ at $p=10$ GeV and $D_s$ with and without inelastic collision channels.
The calculation uses the high likelihood parameter set in Table \ref{table:high-likelihood-parameters}. 
We observe a 30-40\% increase in $\hat{q}$ and similar amount of decrease in $D_s$ if the inelastic contributions are included.

To summarize, we have developed a novel linearized hybrid transport model, called {\tt Lido}, for heavy quark propagation inside a quark-gluon plasma.
Heavy quarks undergo perturbative scatterings with medium particles. Between subsequent scatterings, the propagation is driven by Langevin dynamics with empirical drag and diffusion coefficients.
Model parameters are calibrated using a Bayesian model-to-data analysis by comparing to $D$-meson and $B$-meson observables in Pb+Pb collisions at $\sqrt{s}=5.02$ TeV.
Our results suggest a late onset of medium energy loss.
The diffusion component to the overall transport coefficients is small, with the dominant contribution coming from the explicitly treated scattering processes.

The calibrated model predicts the centrality and $p_T$ dependence of the $B$-meson nuclear modification factor and flows and a non-zero $D$-meson direct flow with respect to $n=3$ event plane.
The extracted heavy quark transport coefficient $\hat{q}$ at finite momenta in the QGP phase is consistent within uncertainties with previous calibrations using an improved Langevin approach as well as with lattice QCD calculations.

\begin{acknowledgments}
  This research was completed using $10^6$ CPU hours provided by the Open Science Grid \cite{Pordes:2007zzb,Sfiligoi:2010zz}, which is supported by the National Science Foundation and the U.S.\ Department of Energy's Office of Science.
SAB, YX, and WK  are supported by the U.S. Department of Energy Grant no. DE-FG02-05ER41367.
 WK is also supported by NSF grant OAC-1550225.
We thank Shanshan Cao, Marlene Nahrgang, and Jussi Auvinen for useful discussion of this project.
\end{acknowledgments}

\begin{appendices}
\section{Running coupling constant}
\label{appendix:alphas}
We use a leading order running coupling constant with three  quark flavors,
\begin{eqnarray}
\alpha_s(Q^2) = \frac{4\pi}{9 \ln\left(Q^2/\Lambda^2\right) }
\end{eqnarray}
The QCD scale is set at $\Lambda = 0.2$~GeV.
Inside a medium, the temperature $T$ defines the medium scale, it is used as a lower cutoff for $Q^2$ of all process, and the running coupling is actually,
\begin{eqnarray}
\alpha_s = \alpha_s(\max\{Q^2,(\mu\pi T)^2\})
\end{eqnarray}
$\mu$ the only parameter we tune in the scattering component of the {\tt Lido} model.
For elastic scattering, $Q^2$ is chosen as the momentum exchange squared for $s,t,u$ channel.
For the gluon emission / absorption vertex, we choose $Q^2 = k_\perp^2$.

\begin{comment}
\section{Calculation of $v_n\{2\}$ and $v_2\{EP\}$}
Cumulant method correlates a heavy meson with a light particle to estimate $v_n$. 
\begin{eqnarray}
v_n\{2\} &=& \frac{d_n\{2\}}{\sqrt{c_n\{2\}} } \\
d_n\{2\} &=& \left\langle \frac{\Re\{pQ^*\}}{mM} \right\rangle_{w = mM} \\
c_n\{2\} &=& \left\langle \frac{|Q|^2-M}{M(M-1)} \right\rangle_{w = M(M-1)}
\end{eqnarray}
\end{comment}

\section{Matrix elements}
\label{appendix:matrix-element}
The vacuum matrix-elements are,
\begin{eqnarray}
\overline{|M_{22,Qq}|^2} &=& \frac{64\pi^2\alpha_s^2}{9} \frac{(M^2-u)^2 + (s-M^2)^2 + 2 M^2 t}{t^2}
\nonumber
\\
\overline{|M_{22,Qg}|^2} &=& \pi^2 \left\{
32\alpha_s^2 \frac{(s-M^2)(M^2-u)}{t^2} \right.
\nonumber
\\
&+&\frac{64}{9}\alpha_s^2 \frac{(s-M^2)(M^2-u)+2M^2(s+M^2)}{(s-M^2)^2} \nonumber
\\
&+&\frac{64}{9}\alpha_s^2 \frac{(s-M^2)(M^2-u)+2M^2(u+M^2)}{(M^2-u)^2} \nonumber
\\
&+& \frac{16}{9}\alpha_s^2 \frac{M^2(4M^2 - t)}{(M^2-u)(s-M^2)} 
\nonumber
\\
&+& 16 \alpha_s^2 \frac{(s-M^2)(M^2-u)+M^2(s-u)}{t(s-M^2)}
\nonumber
\\
&-& \left. 16 \alpha_s^2 \frac{(s-M^2)(M^2-u)-M^2(s-u)}{t(M^2-u)}\right\}
\nonumber
\\
|M_{2\rightarrow 3}|^2 &=& |M_{2\rightarrow 2}|^2 48 \pi \alpha_s (1-\bar{x})^2
\nonumber
\\
&\times&\left(\frac{\vec{k}_\perp}{k_\perp^2 + x^2 M^2} + \frac{\vec{q}_\perp - \vec{k}_\perp}{(\vec{q}_\perp-\vec{k}_\perp)^2 + x^2 M^2}
\right)^2 
\end{eqnarray}
In medium, the denominator of the squared gluon propagator is replaced by $t^2 \rightarrow t(t-m_D^2)$. 
For the radiation processes, we also include a gluon mass to regulate soft divergence $x^2M^2 \rightarrow x^2M^2 + (1-x)m_g^2$, where $m_g^2 = m_D^2/2$ is the squared asymptotic gluon mass. 

\section{Many-body phase space sampling}
\label{appendix:sample}
The phase-space sampling of Equation \ref{eq:rate} is performed sequentially for the initial state and final state phase-space.
For $2\rightarrow 2$ and $2\rightarrow 3$ body processes, we rewrite the integrated rate in the fluid cell rest frame as,
\begin{eqnarray}
\Gamma(E_1, T, t) &=& \frac{d}{\nu} \frac{1}{2E_1}\int \frac{e^{-\beta E_2}dp_2^3}{(2\pi)^32E_2} 
\int d\Phi_m\overline{|M|^2}.
\end{eqnarray}
The nested integration is a Lorentz invariant quantity, and we choose to calculate it in the CoM frame of the collision, 
\begin{eqnarray}
\int d\Phi_m\overline{|M_{22}|^2} &=& 2E_12E_2v_{\textrm{rel}}\sigma \nonumber \\
 &=& 2(s-M^2)\sigma_{\textrm{CoM}}^{22}(\sqrt{s}, T)\nonumber \\
  &=& F_{22},\\
\int d\Phi_m\overline{|M_{23}|^2} &\rightarrow& \int d\Phi_m\overline{|M_{23}|^2} C\left(\frac{\Delta t}{\tau_f}\right) \nonumber \\
 &=& 2(s-M^2)\sigma_{\textrm{CoM}}^{23}(\sqrt{s}, T, \Delta t)\nonumber \\
 &=& F_{23}
\end{eqnarray}
where $\sigma$ is the cross-section of the process.
The phase space integration of the $2\rightarrow 3$ process is modified by the coherence factor $C$ from Equation (\ref{eq:LPM}), so the cross-section is $\Delta t$ dependent.
In practice, the values of the integrated rates and cross-sections are tabulated. 
The sampling of initial state $p_2$ determines the $\sqrt{s}$ of the process, and subsequently we sample the differential cross-section with $\sqrt{s}, T$ (and $\Delta t$) as inputs.

The sampling of $3\rightarrow 2$ body process is more difficult to set up:
the integrated rate is
\begin{eqnarray}
\frac{d}{\nu} \int \frac{e^{-\beta E_2}dp_2^3}{(2\pi)^32E_2} \frac{e^{-\beta k}dk^3}{(2\pi)^32k}C\left(\frac{\Delta t}{\tau_f}\right)
\int d\Phi_2\overline{|M|^2}.
\end{eqnarray}
The Lorentz invariant nested integral is an intricate function of the initial 3-body state kinematics and temperature,
\begin{eqnarray}
\int d\Phi_2\overline{|M|^2} = F_{32}(\sqrt{s}, \sqrt{s_{12}}, \sqrt{s_{1k}}, T).
\end{eqnarray}
Where $s = (p_1+p_2+k)^2$ is the center of mass energy, $s_{12} = (p_1+p_2)^2$ and $s_{1k} = (p_1+k)^2$.
This requires four-dimensional table for the value of $F_{32}$ and a five-dimensional initial state sampling.

The situation would be far more complicated if we utilize quantum statistics or the full HTL propagator, since the former introduces factors like $1\pm f(p\cdot u)$ and the latter introduces to $F_{nm}$ a self energy that depends on the medium rest frame.
In both cases, the Lorentz invariance of $F_{nm}$ is broken and it further depends on $v_{\textrm{CoM}}$, increasing the dimensionality of the problem. 
We are looking for the strategies to include these features in future studies.

\section{Construction of the covariance matrix in the likelihood function}
\label{appendix:sigma}
Construction of the covariance matrix in the likelihood function is not an uniquely defined task.
In principle, the covariance matrix should include theoretical and experimental   uncertainties and the Gaussian process emulator's interpolation uncertainty.
\begin{eqnarray}
\Sigma{ij} = \Sigma^{\textrm{theory}}_{ij} + \Sigma^{\textrm{exp}}_{ij} + \Sigma^{\textrm{emulator}}_{ij}
\end{eqnarray}
For the experimental uncertainty, the statistical errors are uncorrelated and are therefore a diagonal matrix. 
Treatment of the systematic errors is complicated as there could be correlations among them, which are rarely published in the literature. 
In this work, we treat most systematic errors as uncorrelated, except for the correlations between the experimental data points (from the same collaboration) of different centrality bins but with the same $p_T$ bins.
The reason is that for $R_{AA}$ measurements, different centrality bins use the same $p-p$ collision reference and any reference uncertainty should affect all centralities in the same way.
The ansatz for the experimental covariance matrix is, 
\begin{eqnarray}
\Sigma^{\textrm{exp}}_{ij} &=& \delta_{ij}\left(\sigma_i^2\right)^{\textrm{stat, uncorr sys}} \nonumber\\
&+& C \left(\sigma_{i}\sigma_{j}\right)^{\textrm{corr sys}}
\end{eqnarray}
The last term is constructed for the correlations over centrality where the prefect correlation matrix is reduced by a factor $C=0.6$.
For the theoretical uncertainty estimation, an additional diagonal uncertainty is introduced with variable magnitude $\sigma^{\textrm{model}}$,
\begin{eqnarray}
\Sigma^{\textrm{theory}}_{ij} = \delta_{ij}(\sigma^{\textrm{model}})^2.
\end{eqnarray}
The $\sigma^{\textrm{model}}$ parameter was given a gamma-distribution prior and is marginalized in the MCMC process.
For the Gaussian process emulator uncertainty, we simply use the predicted variance at each point.
\end{appendices}
\bibliography{hybrid} 
\end{document}